\documentclass{article}
\usepackage[dvips]{graphicx}

\begin{document}
\title{New Index of CP Phase Effect and $\theta_{13}$ Screening \\
in Long Baseline Neutrino Experiments}
\author{
\sc
{Keiichi Kimura$^1$}
{Akira Takamura$^{1,2}$} \\
\sc
{and} \\
\sc
{Tadashi Yoshikawa$^1$}
\\
\\
{\small \it $^1$Department of Physics, Nagoya University,}
{\small \it Nagoya, 464-8602, Japan}\\
{\small \it $^2$Department of Mathematics,
Toyota National College of Technology}\\
{\small \it Eisei-cho 2-1, Toyota-shi, 471-8525, Japan}}
\date{}
\maketitle

\vspace{-9.5cm}
\begin{flushright}
\end{flushright}
\vspace{7.5cm}

\begin{abstract}
We introduce a new index of the leptonic CP phase dependence
$I_{\rm CP}$ and
derive the maximal condition for this index in a simple and
general form.
$I_{\rm CP}\simeq 100\%$ may be realized even in the JPARC experiment.
In the case that the 1-3 mixing angle can be observed in the next
generation reactor experiments, namely $\sin^2 2\theta_{13}>0.01$,
and nevertheless $\nu_e$ appearance signal cannot be observed
in the JPARC experiment,
we conclude that the CP phase $\delta$ becomes
a value around $135^{\circ}$
$(45^{\circ})$ for $\Delta m^2_{31}>0$ $(\Delta m^2_{31}<0)$ 
without depending the uncertainties of solar and atmospheric 
parameters.
\end{abstract}


\maketitle


\section{Introduction}

In future experiments, the determination of the leptonic CP phase
$\delta$ is one of the most important aim in elementary particle
physics.
A lot of effort have been dedicated both from theoretical and
experimental point of view in order to attain this aim,
see \cite{Arafune9703, Cervera, Minakata0402, Huber}
and the references therein.

The CP asymmetry,
$A_{\rm CP}=(P_{\mu e}-\bar{P}_{\mu e})/(P_{\mu e}+\bar{P}_{\mu e})$,
is widely used as the index of the CP phase dependence.
Here, $P_{\mu e}$ and $\bar{P}_{\mu e}$ are
the oscillation probabilities for the transition
$\nu_{\mu} \to \nu_e$ and $\bar{\nu}_{\mu} \to \bar{\nu}_e$
respectively.
However, this index has to be improved on the following three points.
The first one is that the fake CP violation due to matter effect
\cite{MSW} cannot be separated clearly in $A_{\rm CP}$.
The second one is that only the effect originated from $\sin \delta$
is included in $A_{\rm CP}$.
The third one is that we need to observe the channels both in neutrino
and anti-neutrino for calculating $A_{\rm CP}$.

In this letter, we introduce a new index of the CP phase dependence
improved on the above three points.
In arbitrary matter profile, we derive the maximal condition of this index
exactly for $\nu_{\mu} \to \nu_e$ transition.
This index can be extended to the case for other channels and
other parameters \cite{future}.
We can simply find the situation that the CP phase effect becomes large
by using this index.
As an example, we demonstrate the following interesting phenomena.
It is commonly expected that a large $\nu_e$ appearance signal
is observed in the JPARC experiment
\cite{JHF} if the 1-3 mixing angle $\theta_{13}$ is relatively large
$\sin^2 2\theta_{13}>0.01$ and is determined
by the next generation reactor experiments like the Double Chooz experiment
\cite{double chooz} and the KASKA experiment \cite{kaska}.
However, there is the possibility that $\nu_e$ appearance signal cannot
be observed
in certain mass squared differences and mixing angles even the case
for large $\theta_{13}$. We call this ``$\theta_{13}$ screening''.
This occurs due to the almost complete cancellation of the large
$\theta_{13}$ effect by the CP phase effect.
If the background can be estimated precisely,
we can obtain the information on the CP phase through the $\theta_{13}$
screening.
This means that we cannot neglect the CP phase effect, 
which is actually neglected in many investigations as the first
approximation.

\section{General Formulation for Maximal CP Phase Effect}

At first, we write the Hamiltonian in matter \cite{MNS} as
\begin{equation}
H = O_{23} \Gamma H'\Gamma^\dagger O_{23}^T
\end{equation}
by factoring out the 2-3 mixing angle and the CP phase,
where $O_{23}$ is the rotation matrix in the 2-3 generations and
$\Gamma$ is the phase matrix defined by
$\Gamma={\rm diag}(1,1,e^{i\delta})$.
The reduced Hamiltonian $H'$ is given by
\begin{eqnarray}
H'=O_{13}O_{12}{\rm diag}(0,\Delta_{21},\Delta_{31})O_{12}^TO_{13}^T
+ {\rm diag}(a,0,0),
\end{eqnarray}
where $\Delta_{ij}=\Delta m_{ij}^2/(2E)=(m_i^2-m_j^2)/(2E)$,
$a=\sqrt2 G_F N_e$, $G_F$ is the Fermi constant, $N_e$ is the electron
number density, $E$ is neutrino energy and $m_i$ is the mass of $\nu_i$.
The oscillation probability for $\nu_{\mu} \to \nu_e$ is proportional
to the $\cos \delta$ and $\sin \delta$ in arbitrary matter profile
\cite{Kimura0203}
and can be expressed as
\begin{eqnarray}
P_{\mu e}=A\cos \delta+B\sin \delta+C
=\sqrt{A^2+B^2}\sin (\delta+\alpha)+C \label{4}.
\end{eqnarray}
Here $A$, $B$ and $C$ are determined by parameters other than
$\delta$ and are calculated by
\begin{eqnarray}
A&=&2{\rm Re}[S_{\mu e}^{\prime *}S_{\tau e}^{\prime}]c_{23}s_{23}
\label{eq A}, \\
B&=&2{\rm Im}[S_{\mu e}^{\prime *}S_{\tau e}^{\prime}]c_{23}s_{23}, \\
C&=&|S_{\mu e}^{\prime}|^2c_{23}^2+|S_{\tau e}^{\prime}|^2s_{23}^2
\label{eq C},
\end{eqnarray}
where $S_{\alpha\beta}^{\prime}=\left[\exp (-iH' L) \right]_{\alpha\beta}$,
$\tan \alpha=A/B$ and $\sqrt{A^2+B^2}\sin (\delta+\alpha)$ is the
CP dependent term and $C$ is the CP independent term.

Next, let us introduce a new index of the CP phase dependence $I_{\rm CP}$.
Suppose that $P_{max}$ and $P_{min}$ as the maximal and minimal values
when $\delta$ changes from $0^{\circ}$ to $360^{\circ}$.
Then, we define $I_{\rm CP}$ as
\begin{equation}
I_{\rm CP}=\frac{P_{max}-P_{min}}{P_{max}+P_{min}}=\frac{\sqrt{A^2+B^2}}{C}.
\end{equation}
Namely, the new index is expressed by the ratio of the coefficient of
the CP dependent term to the CP independent term.
$I_{\rm CP}$ is a useful tool to explore where is the most effective
region in parameter spaces to extract the CP effect from long baseline
experiments although $I_{\rm CP}$ is not an observable.
$A_{\rm CP}$ is also similar one and this is an observable. 
However $A_{\rm CP}$ have to be expressed by $\delta$ 
though $\delta$ is still unknown parameter so that
$A_{\rm CP}$ seems not to be so good index to make the exploration.
On the other hand, $I_{\rm CP}$ is calculated
without using $\delta$.
This is the main difference between these two indices 
and it is more effective to use $I_{\rm CP}$.
\begin{center}
\begin{tabular}{cc}
    \resizebox{64mm}{!}{\includegraphics{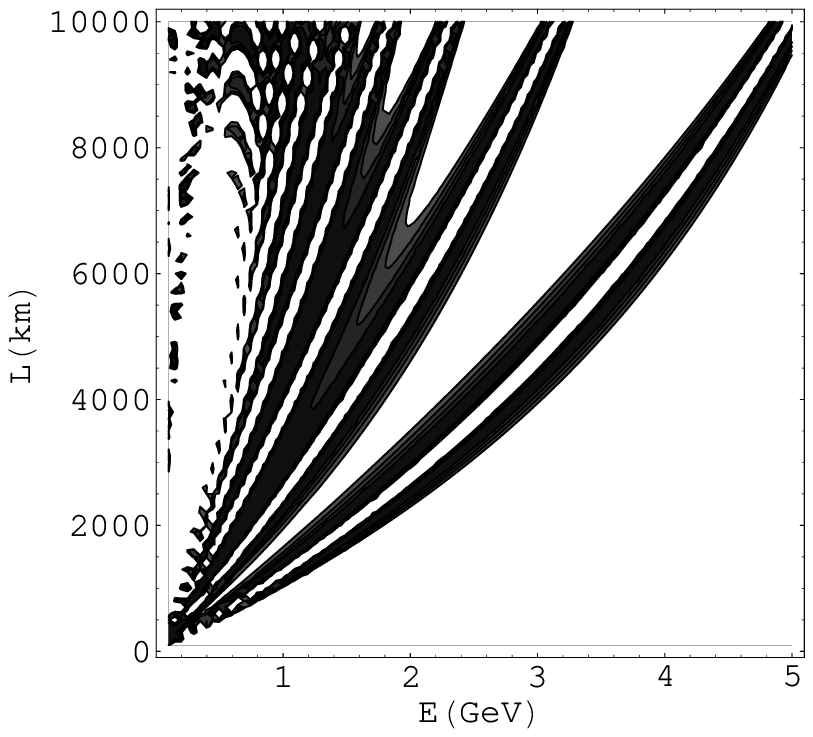}} &
    \resizebox{64mm}{!}{\includegraphics{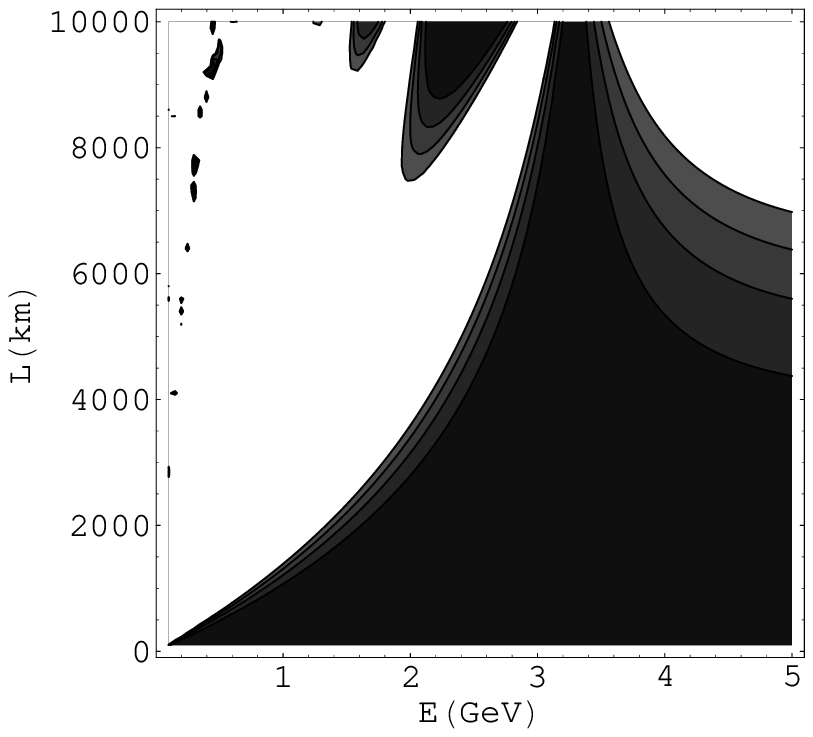}}
\end{tabular}

\vspace{-0.2cm}
\begin{flushleft}
Fig.1. Region with large $I_{\rm CP}$.
We write the lines for $I_{\rm CP}=95\%$, $90\%$, $85\%$, $80\%$.
$I_{\rm CP}$ takes the large value in black region.
Left and right panels are drawn with two different sets of
parameters.
\end{flushleft}
\end{center}
In this letter, we show the region for taking the large value of $I_{\rm 
CP}$
in the $E$-$L$ plane.
In particular, we investigate how the region changes by the uncertainties of
atmospheric and solar parameters if $\theta_{13}$ is determined
by the future reactor experiment.
In fig.1(left), we choose the parameters as $\Delta m_{21}^2=7.9\times
10^{-5}
{\rm eV}^2$,
$\sin^2 \theta_{12}=0.31$, $\Delta m_{31}^2=2.2\times 10^{-3} {\rm eV}^2$,
$\sin^2 \theta_{23}=0.50$, which are the best-fit values in
the present experiments \cite{Schwetz0510}.
We also use $\sin^2 2\theta_{13}=0.1$, which is near on the upper bound
of the CHOOZ experiment \cite{CHOOZ}.
On the other hand, in fig.1(right), we choose $\Delta m_{21}^2=8.9\times
10^{-5}{\rm eV}^2$, $\sin^2 \theta_{12}=0.40$,
$\Delta m_{31}^2=1.4\times 10^{-3}{\rm eV}^2$,
$\sin^2 \theta_{23}=0.34$, $\sin^2 2\theta_{13}=0.01$,
which are within the 3-$\sigma$ allowed region respectively
\cite{Schwetz0510}.
We use the $\rho=2.8$g/cm$^3$ as the matter density.
Fig.1(left) shows that $I_{\rm CP}$ takes a large value along the line
$L/E=const.$ in low energy region.
In contrary, fig.1(right) shows that there is a wide region with almost
$I_{\rm CP}\simeq 100\%$.
The region appears in the baseline shorter than $3000$km
and surprisingly it is independent of neutrino energy.
So, what is the condition for realizing the maximal $I_{\rm CP}$ ?
If we notice (\ref{eq A})-(\ref{eq C}),
we find that the relation between the denominator and the numerator of
$I_{\rm CP}$.
Namely, the fact that the average of two positive quantities is in general
larger than the square root of their product yields
\begin{equation}
C=|S_{\mu e}^{\prime}|^2c_{23}^2+|S_{\tau e}^{\prime}|^2s_{23}^2
\geq 2|S_{\mu e}^{\prime *}S_{\tau e}^{\prime}|c_{23}s_{23}=\sqrt{A^2+B^2}.
\end{equation}
In relation to this, Burguet {\it et al.} have pointed out the fact
``the CP dependent term cannot be larger than the CP independent term''
by using the approximate formula \cite{Burguet0103}.
In this letter, we define $I_{\rm CP}$ without depending on the unknown CP
phase and derive the exact inequality in arbitrary matter profile
for the first time.
Furthermore, we consider the condition that both sides become equal
in this inequality.
This condition is given by
\begin{eqnarray}
|S_{\mu e}^{\prime}|c_{23}=|S_{\tau e}^{\prime}|s_{23}\quad
{\rm (maximal \,\,condition)}
\label{max-condition},
\end{eqnarray}
and $I_{\rm CP}=100\%$ is realized when this condition is satisfied.
Below, let us investigate the maximal condition (\ref{max-condition})
in detail by using the approximate formula
including the non-perturbative effect of small parameters $\theta_{13}$ and
$\Delta m^2_{21}$, in constant matter profile \cite{Takamura0403}.
The reduced amplitudes in the maximal condition are approximated by
\begin{eqnarray}
S^{\prime}_{\mu e}&\simeq& \lim_{s_{13}\to 0}\left[\exp (-iH' L)
\right]_{\mu e}, \\
S^{\prime}_{\tau e}&\simeq&\lim_{\Delta_{21}\to 0}
\left[\exp (-iH' L) \right]_{\tau e}.
\end{eqnarray}
The maximal condition is rewritten as
\begin{equation}
\frac{\Delta_{21}\sin 2\theta_{12}}{\Delta_{\ell}}c_{23}
\sin \left(\frac{\Delta_{\ell}L}{2}\right)
=\left|\frac{\Delta_{31}\sin 2\theta_{13}}{\Delta_{h}}s_{23}
\sin \left(\frac{\Delta_{h}L}{2}\right)\right| \label{max-condition2}
\end{equation}
in this approximation, where $\Delta_h=\Delta m^2_h/(2E)$,
$\Delta_{\ell}=\Delta m^2_{\ell}/(2E)$ and
$\Delta m^2_h$ ($\Delta m^2_{\ell}$) stands for the effective mass
squared differences corresponding to high (low) energy MSW effect.
The concrete expression for $\Delta_h$ is given by
\begin{eqnarray}
\Delta_{h}=\sqrt{\left(\Delta_{31}\cos 2\theta_{13}-a\right)^2
+\Delta_{31}^2\sin^2 2\theta_{13}} \label{15}.
\end{eqnarray}
We obtain $\Delta_{\ell}$ by the replacements
$\Delta_{h} \to \Delta_{\ell}$,
$\Delta_{31} \to \Delta_{21}$,
$\theta_{13} \to \theta_{12}$.
The energy dependence of
$\Delta m^2_h$ and $\Delta m^2_{\ell}$ is mild
and roughly speaking we regard these as constant.
At this time, (\ref{max-condition2}) becomes the equation
for $L/E$ and the region for large $I_{\rm CP}$ appears
along the line $L/E=const.$ in fig.1(left).
On the other hand, there appears no $L/E$ dependence at short
baseline in fig.1(right).
This is interpreted as follows.
In the case of small $x$, the approximation $\sin x\simeq x$ becomes good
and the $E$ and $L$ dependencies of both sides vanish and the maximal
condition can be simplified as
\begin{eqnarray}
\Delta m^2_{21}\sin 2\theta_{12}c_{23}=|\Delta m^2_{31}|\sin
2\theta_{13}s_{23}
\label{screening-condition}
\end{eqnarray}
The inequality for $L$ is obtained by $\Delta_{\ell}L/2\ll 1$ as
\begin{eqnarray}
L\ll\frac{2}{a}=\frac{2}{\sqrt{2}G_F \rho Y_e}
\simeq 3500[{\rm km}],
\end{eqnarray}
where we use $\rho=2.8$g/cm$^3$ as the matter density and $Y_e=0.494$
as the electron fraction.
The inequality for $E$ is also obtained by $\Delta_{h}L/2\ll 1$ as
\begin{eqnarray}
E\gg \frac{\Delta m^2_{31}L}{4}\simeq
\frac{L}{500}[{\rm GeV}],
\end{eqnarray}
where the baseline length $L$ is measured in the unit of km.
The region for satisfying these conditions coincides with that
for taking large $I_{\rm CP}$ in fig.1(right).

Next, let us investigate the condition (\ref{screening-condition}).
That is rewritten as
\begin{equation}
\sin 2\theta_{13}=0.036\times \frac{\Delta m_{21}^2}{8\cdot 10^{-5}}
\frac{\sin 2\theta_{12}}{0.9}\frac{1.0}{\tan \theta_{23}}
\frac{2\cdot 10^{-3}}{|\Delta m_{31}^2|}. \label{18}
\end{equation}
In the case that the parameters except $\theta_{13}$ have their
best-fit values in the present experiments, the value of $\theta_{13}$
satisfying (\ref{18}) becomes small compared with the bound
$\sin 2\theta_{13}>0.1$ of next generation reactor experiments.
However, it is possible to satisfy Eq.(\ref{screening-condition})
if the parameters possess values, which slightly deviate
from those of the best-fit.
We chose the parameters for satisfying Eq.(\ref{screening-condition})
in fig.1(right).
The readers may think that
such a situation with large $I_{\rm CP}$
independent of $E$ and $L$ is extraordinary and
it is not likely realized.
However, if we write
the ratio of both sides of (\ref{screening-condition}) as
\begin{eqnarray}
r=\frac{\Delta m^2_{21}c_{23}\sin 2\theta_{12}}{|\Delta m^2_{31}|s_{23}\sin
2\theta_{13}},
\end{eqnarray}
$I_{\rm CP}$ is calculated by using $r$ as
\begin{equation}
I_{\rm CP}=\frac{2r}{1+r^2},
\end{equation}
and we found that the large CP phase effect is realized
as $I_{\rm CP}=97.5\% (88\%)$ with $r=0.8(0.6)$ for example.
Thus, the decrease of $I_{\rm CP}$ according to the
difference from Eq.(\ref{screening-condition}) is comparatively mild
and the region with large $I_{\rm CP}$ becomes wider than expected.
If we describe for the reference, $r=0.10 (0.87)$ and
$I_{\rm CP}\simeq 20\% (99\%)$ in fig.1(left) (fig.1(right)).
The discovery of such a situation, where the maximal condition is
satisfied
independently of $E$ and $L$ is one of our main results in this letter.
This may be realized in the JPARC experiment
under the condition of certain parameter combinations,
and is very important to analyze the experimental result.

\section{$\theta_{13}$ Screening in JPARC Experiment}

We found that the screening condition may be realized in the
JPARC experiment.
Next, we consider the case such that $I_{\rm CP}$ takes
a large value and the CP phase effect contributes to the probability
destructively.
It is commonly expected that a large $\nu_e$ appearance signal
is observed
in the JPARC experiment, if $\theta_{13}$ is large and
will be confirmed in next generation reactor experiments.
We point out here that the probability can be zero over the
entire region in the JPARC experiment due to the cancellation
of $\theta_{13}$ effect by the CP phase effect.
We use the same parameters as in fig.1(right), where
Eq.(\ref{screening-condition}) holds.
Fig.2 shows the oscillation probabilities with
$\delta=45^{\circ}, 135^{\circ}, 225^{\circ}, 315^{\circ}$
in the energy range $0.4$-$1.2$ GeV of the JPARC experiment.
In fig.2, one can see that the CP dependent term has the same sign
as the CP independent term when $\delta=315^{\circ}$.
They interfere constructively with each other and generate the large
probability.
On the other hand, they have opposite sign and almost completely cancel
each other when $\delta=135^{\circ}$.
As a result, the probability for $\nu_{\mu} \to \nu_e$ transition
is strongly suppressed and we call this phenomena ``$\theta_{13}$
screening''.

\begin{center}
\begin{tabular}{l}
\hspace{-1cm}
\resizebox{90mm}{!}{\includegraphics{probability.eps}}
\end{tabular}
 \vspace{-0.2cm}
\begin{flushleft}
Fig.2. CP dependence of $P_{\mu e}$ under the maximal condition.
Four lines stand for the oscillation probabilities
with $\delta=45^{\circ}, 135^{\circ}, 225^{\circ}$ and $315^{\circ}$,
respectively.
\end{flushleft}
\end{center}

Let us calculate the value of $\delta$ for the $\theta_{13}$ screening.
At first, the value of $\alpha$ is determined by $\sin
\alpha=A/\sqrt{A^2+B^2}$,
$\cos \alpha=B/\sqrt{A^2+B^2}$.
This leads to
\begin{eqnarray}
\tan \alpha=\frac{A}{B}\simeq
\frac{-1}{\tan \left(\frac{\Delta m^2_{32}L}{4E}\right)}
=\tan\left(\frac{\pi}{2}+\frac{\Delta m^2_{32}L}{4E}\right)
\label{21}.
\end{eqnarray}
Substituting $\Delta m^2_{32}=1.4\times 10^{-3} {\rm eV}^2$,
$L=295$km and $E=0.7$GeV into this relation, we obtain
$\alpha\simeq 135^{\circ}$.
If $\delta\simeq 135^{\circ}$, we obtain
$\sin (\delta+\alpha)=\sin 270^{\circ}=-1$
and as a result $C$ and $\sqrt{A^2+B^2}$ cancel each other
from eq. (\ref{4}).
As seen from (\ref{21}), the value of $\delta$ for the $\theta_{13}$
screening
changes with $E$ and $L$ in general.
However, $\sin (\delta+\alpha)$ takes a local minimum around
$\delta+\alpha=270^{\circ}$ and the magnitude of the CP
dependent term changes at most $10\%$ even if $\alpha$ changes
$30^{\circ}$ around this minimum.
This is the reason for $P_{\mu e}\simeq 0$ in a wide energy region.

Here, we discuss the relation between the problems of parameter degeneracy
and the $\theta_{13}$ screening.
In this letter, we investigated the case for large $\theta_{13}$,
which will be confirmed by next generation reactor experiments.
Namely, we have considered the case of $\theta_{13}$-$\delta$ ambiguity free
\cite{Burguet0103}.
Next, let us consider the $\theta_{23}$ ambiguity \cite{Barger}.
In order for the $\theta_{13}$ screening to be realized,
the parameters should satisfy the relation (\ref{screening-condition}).
Under this condition, only a small $\theta_{23}$ near the present lower
bound
is permitted, namely $\theta_{23}\simeq 35^{\circ}$,
which corresponds to $\sin^2 \theta_{23}=0.34$.
Therefore, if the $\theta_{13}$ screening is observed,
$\theta_{23}$ degeneracy is solved.
Finally, let us consider the $\Delta m^2_{31}$ sign ambiguity
\cite{Minakata}.
In the case for also $\Delta m^2_{31}<0$, the maximal condition
is almost independent of $E$ and $L$ at the baseline of the JPARC
experiment $L=295$ km.
The sign of $A$ changes and becomes negative according to the
replacement of the sign of $\Delta m^2_{31}$.
On the other hand, the sign of $B$ does not change and is negative.
This leads to $\alpha=225^{\circ}$ and the $\theta_{13}$ screening
occurs around $\delta=45^{\circ}$ for $\Delta m^2_{31}<0$.

Next, in fig.3, we numerically estimate
$\nu_e$ appearance signal, namely the total number of events
distinct from the background noise,
obtained in the JPARC-SK experiment within the energy range $E=0.4$-$1.2$GeV
when the CP phase $\delta$ changes from $0^{\circ}$ to $360^{\circ}$.
In top and down figures, we use $\Delta m_{31}^2=1.4\times 10^{-3}{\rm 
eV}^2$
and $3.0\times 10^{-3}{\rm eV}^2$ respectively.
In left and right figures, we use $\sin^2 \theta_{23}=0.34$ and $0.66$.
Other parameters are taken as in fig.1 (right).
We assume here only the neutrino beam data as realized in
the JPARC-SK for five years.
We also give the statistical error within the 1-$\sigma$ level in fig.3.
We use the globes software to perform the numerical calculation
\cite{globes,JHF}.

\begin{center}
\begin{tabular}{cc}
  \resizebox{74mm}{!}{\includegraphics{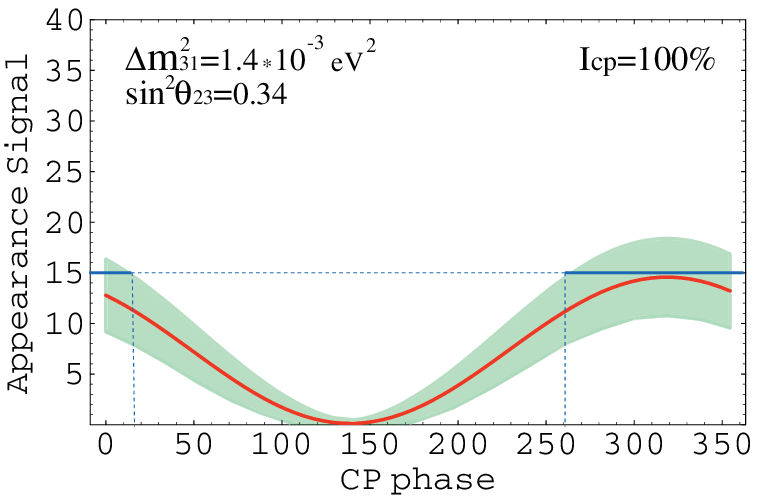}} &
  \resizebox{74mm}{!}{\includegraphics{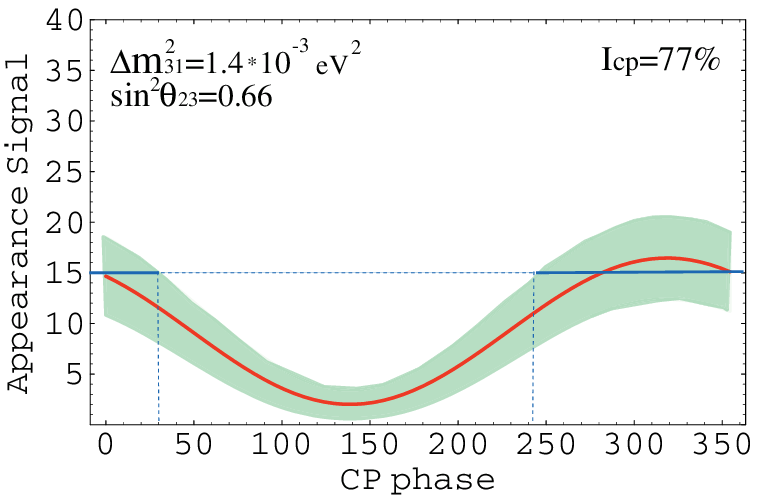}} \\
  \resizebox{74mm}{!}{\includegraphics{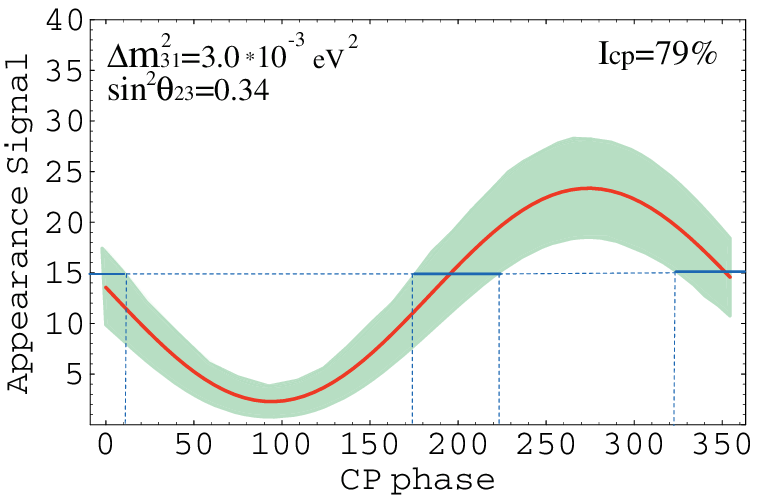}} &
  \resizebox{74mm}{!}{\includegraphics{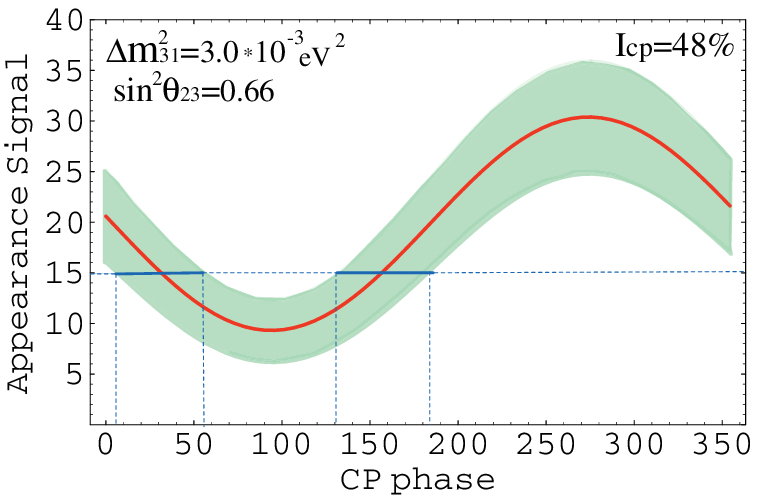}} \\
\end{tabular}

\vspace{-0.2cm}
\begin{flushleft}
Fig.3. CP dependence of $\nu_e$ appearance signal in the JPARC-SK 
experiment.
We use the parameters as in fig.1 (right) except for $\Delta m_{31}^2$ and
$\sin^2 \theta_{23}$.
The condition (\ref{screening-condition}) is satisfied in the top-left figure,
and is not satisfied in other figures.
The statistical error is also shown within the 1-$\sigma$ level.
We also show the value of $I_{\rm CP}$ calculated at $1$GeV in figures.
\end{flushleft}
\end{center}

As expected from the oscillation probability in fig.2,
$\nu_e$ appearance signal will become almost zero around 
$\delta=135^{\circ}$
even during the five years data acquisition in the SK experiment
in the top-left of fig. 3.
Note that this occurs only when the maximal condition
(\ref{screening-condition}) is satisfied, namely $I_{\rm CP}\simeq 100\%$.
Other panels in fig.3 show that the minimal value of $\nu_e$ appearance 
signal
rise and is a little different from zero
because (\ref{screening-condition}) is not satisfied so precisely.
We obtain the similar results in the case that
$\Delta m^2_{21}$ or $\sin^2 \theta_{12}$ changes within the allowed region
obtained from solar and the KamLAND experiments.
Let us here illustrate how the value of $\delta$ is constrained
by the experiment.
Below, suppose that the atmospheric
parameters have some uncertainties as $\Delta m^2_{31}=1.4$-
$3.0\times 10^{-3}{\rm eV}^2$ and $\sin^2 \theta_{23}=0.34$-$0.66$,
while the solar parameters $\Delta m^2_{21}$ and $\sin^2 \theta_{12}$
and $\sin^2 \theta_{13}$ are fixed for simplicity.
For example, if 15 appearance signals are observed in the experiment,
we obtain the allowed region of $\delta$ as $0^{\circ}$-$50^{\circ}$
or $130^{\circ}$-$220^{\circ}$ or $240^{\circ}$-$360^{\circ}$
from four figures in fig.3.
Namely, combined range of all allowed region is totally $260^{\circ}$.
Next, we consider the case that no appearance signal is obtained.
This gives the allowed region of $\delta$ as $110^{\circ}$-$160^{\circ}$.
Namely, combined range of all allowed region is totally $50^{\circ}$.
Thus, we found from above rough estimation
that the stronger constraint is obtained
in the case of $\theta_{13}$ screening
even if the uncertainties of parameters except for $\delta$ are considered.
In other words, we can also obtain the information on the atmospheric and
solar parameters.
Although the precise estimation of the background is a difficult problem,
it is interesting to have strong constraint for
not only the value of the CP phase but also other parameters like
$\Delta m_{31}^2$ and $\sin^2 \theta_{23}$
when the $\nu_e$ appearance signal is not observed
and the 1-3 mixing angle has a comparatively large value
$\sin^2 2\theta_{13}>0.01$.

\section{Summary and Discussion}

 \label{sec:summary}
In summary, we introduced a new index of the
CP phase dependence $I_{\rm CP}$ and derived their maximal condition
in a simple and general form.
In particular, we showed that $I_{\rm CP}\simeq 100\%$ is
realized in a rather wide region in the $E$-$L$ plane at certain values
of parameters.
In the case that $\theta_{13}$ has a comparatively large value
$\sin^2 2\theta_{13}=0.01$, (namely $\theta_{13}$ will be
observed in next generation reactor experiments)
nevertheless we cannot observe $\nu_e$ appearance signal in the JPARC
experiment,
we obtain the information on the CP phase as
$\delta\simeq 135^{\circ}\, (\delta\simeq 45^{\circ})$
for $\Delta m^2_{31}>0\,(\Delta m^2_{31}<0)$
without depending on the uncertainties of other parameters.
Also for $\sin^2 2\theta_{13}<0.01$, there is a possibility
that the $\theta_{13}$ screening will occur.
In this case, we need to consider the reason for the absence of $\nu_e$
appearance signal in the JPARC experiment more carefully,
in order to understand whether $\theta_{13}$ is small or the $\theta_{13}$
effect is canceled out by the CP phase effect.
We also note that the $\theta_{13}$ screening may be realized for
not only $\nu_{\mu} \to \nu_e$ oscillation in super-beam experiments
but also $\nu_e \to \nu_{\tau}$ oscillation in neutrino factory
experiments.
We can also use the zero probability in the $\theta_{13}$ screening
to explore new physics like non-standard interaction.

\section*{Acknowledgement}

We would like to thank Prof. Wilfried Wunderlich
(Tokai university) for helpful comments and
advice on English expressions.

\end{document}